# Magnetocaloric properties of a frustrated Blume-Capel antiferromagnet

M. Žukovič and A. Bobák

Department of Theoretical Physics and Astrophysics, Institute of Physics, P.J. Šafárik University in Košice, Park Angelinum 9, 041 54 Košice, Slovakia

**Abstract.** Low-temperature magnetization processes and magnetocaloric properties of a geometrically frustrated spin-1 Blume-Capel model on a triangular lattice are studied by Monte Carlo simulations. The model is found to display qualitatively different behavior depending on the sign of the single-ion anisotropy $D$. For positive values of $D$ we observe two magnetization plateaus, similar to the spin-1/2 Ising antiferromagnet, and negative isothermal entropy changes for any field intensity. For a range of small negative values of $D$ there are four magnetization plateaus and the entropy changes can be either negative or positive, depending on the field. If $D$ is negative but large in absolute value then the entropy changes are solely positive.

## 1 Introduction

Investigations of the magnetocaloric effect (MCE) have greatly intensified since the discovery of the giant magnetocaloric effect in $Gd_5(Si_xGe_{1-x})_4$ [1]. MCE is characterized by the isothermal magnetic entropy change $\Delta S$, when an external magnetic field is varied. Depending on the sign of the temperature derivative of the magnetization, it can be either negative (a conventional MCE observed in regular ferromagnetic materials) or positive (an inverse MCE [2]). In the former case the sample heats up when the external magnetic field is applied adiabatically, while in the latter case it cools down. Theoretic description of the inverse MCE (IMCE) in antiferromagnetic and ferrimagnetic systems was recently provided within mean-field approximation [3]. IMCE in antiferromagnetic systems is associated with antiparallel arrangement of magnetic sublattices, which due to thermal fluctuations tend to align with the direction of the magnetic field, thus causing increase of the total magnetization with temperature. IMCE has also been observed in real antiferromagnetic compounds, such as $MnBr_2 \cdot 4H_2O$ [4].

It is known, however, that frustrated magnetic systems may show magnetization processes quite different from their nonfrustrated counterparts and, therefore, one can also expect different magnetocaloric behavior. Some theoretical studies have predicted in both classical [5] and quantum [6] frustrated Heisenberg antiferromagnets MCE that is enhanced compared with the nonfrustrated systems. Improved magnetocaloric properties due to frustration have also been reported in geometrically frustrated compounds $Gd_2Ti_2O_7$ [7] and $TbNiAl$ [8]. Recently, a Monte Carlo study [9] has predicted anomalous magnetocaloric phenomena related to the field-induced magnetization plateaus in the geometrically frustrated classical spin-1/2 triangular lattice Ising antiferromagnet (TLIA).

In the present study we employ Monte Carlo simulations in order to study magnetization processes and magnetocaloric properties of a generalized spin-1 TLIA model, involving a single-ion anisotropy term which is expected to play a significant role in the magnetocaloric behavior of the system.

## 2 Method

We consider an antiferromagnetic spin-1 Blume-Capel (BC) model on a triangular lattice in an external magnetic field $h$, described by the Hamiltonian

$$H = -J \sum_{<i,j>} S_i S_j - D \sum_i S_i^2 - h \sum_i S_i, \qquad (1)$$

where $J < 0$ is an antiferromagnetic exchange interaction, $D$ is a single-ion anisotropy parameter and $S_i = \pm 1, 0$. We employ standard Monte Carlo simulations with Metropolis dynamics and periodic boundary conditions. Since magnetization processes are little sensitive to the system size (as also observed in Ref. [9]), a relatively small linear lattice size of $L = 24$ was used in the simulations throughout this study. Consequently, the equilibration was relatively fast and $10^4$ MC sweeps were sufficient to bring the system to the equilibrium. Then, for thermal averaging we used another $5 \times 10^4$ MC sweeps. In order to obtain reduced field $h/|J|$ dependencies at fixed values of reduced temperature $k_B T/|J| = 0.1$ and several selected values of the reduced single-ion anisotropy $D/|J|$ we proceed as follows. Simulations start from $h/|J| = 0$ using an appropriate initial state, such as those indicated in figure 1. Then the field is gradually increased with the step $\Delta h/|J| = 0.01$ and the simulations start from the final configuration obtained at the previous field value. We evaluate the total magnetization per spin

$$m = \langle M \rangle / L^2 = \left\langle \sum_i S_i \right\rangle / L^2, \qquad (2)$$

and the isothermal magnetic entropy change per spin $\Delta s$, which occurs at changing the field from 0 to $h$, is estimated from thermodynamic Maxwell equation (with $k_B = 1$) by numerical quadrature

$$\Delta s(T,h) = \int_0^h \left(\frac{\partial m}{\partial T}\right)_h dh. \qquad (3)$$

## 3 Results

### 3.1. Ground states

Reduced zero-temperature energies per spin of different configurations can be obtained from the Hamiltonian (1) in the form

$$e/|J| = \sum_{\langle k,l \rangle} S_k S_l - \frac{D}{3|J|} \sum_k S_k^2 - \frac{h}{3|J|} \sum_k S_k, \qquad (4)$$

where the first summation runs over the nearest neighbors $S_k$ and $S_l$ ($k, l = 1, 2, 3$) on the triangular plaquette formed by neighboring spins. Then, the ground states in the model parameter space $h/|J|$-$D/|J|$, presented in figure 1, can be determined from the minimum energy condition. The phase boundaries are obtained by equating expressions for energies in various phases. In total we obtained seven different phases, corresponding to different sublattice magnetization arrangements. We note that, except for the nonmagnetic phase (○○○) at low fields when all the spins are in state $S_i = 0$ and saturated phase (↑↑↑) at high fields when all the spins are fully aligned with the field direction, all the remaining configurations are degenerate as follows: (○○↑), (○↑↑) and (↑↓↑) are three-times and (○↓↑) and (↓↓↑) are six-times degenerate.

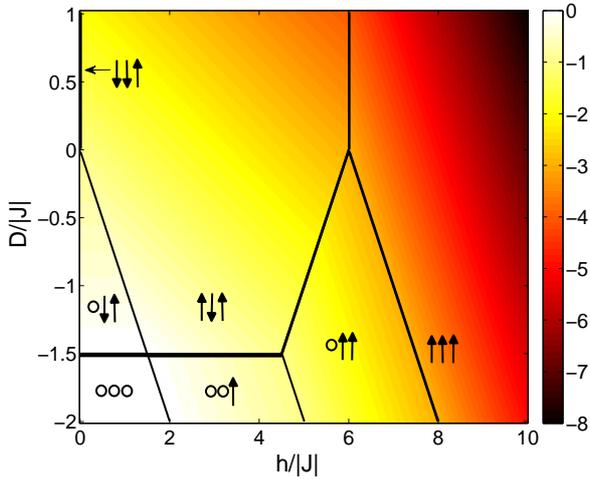

**Fig. 1.** Ground state energy in $h/|J|$-$D/|J|$ plane. Solid lines separate different phases with different sublattice magnetizations arrangements: 0 (○), +1 (↑) and -1 (↓).

### 3.2. Low-temperature behavior

Based on the above ground-state considerations, apparently, the low-temperature field-increasing magnetization processes will strongly depend on the single-ion anisotropy strength $D$. Negative values of $D$ tend to enhance nonmagnetic states ($S_i = 0$) and thus the model is expected to behave like a magnetically diluted spin system. On the other hand, positive values of $D$ tend to enhance magnetic states ($S_i = \pm 1$) and thus the situation much resembles the spin-1/2 case. Namely, as one can see in figure 2, at $k_B T/|J| = 0.1$ for $D/|J| = 0$ and 1 in a field there is one plateau in the ferrimagnetic phase (↑↓↑) with the total magnetization equal to one third of the saturation value, followed by the paramagnetic phase (↑↑↑) with the fully saturated magnetization. The spin-1/2 TLIA case is also included for comparison. Both the spin-1 case with $D/|J| \geq 0$ and the spin-1/2 case show similar magnetization processes, taking into account the fact that in the ground state the plateau heights and the transition field values of the latter are half of those of the former. However, for $-3/2 < D/|J| < 0$, besides the one third magnetization plateau, there are two additional plateaus. At low fields there is a plateau with zero magnetization (○↓↑) and before the saturation phase there is another plateau with the height of two thirds of the saturation value (○↑↑). For $D/|J| < -3/2$, the first two plateaus are again of the same heights of zero and one third of the saturation value but, as indicated in figure 1, they result from different sublattice magnetizations, (○○○) and (○○↑), respectively. Nevertheless, the magnetocaloric properties are qualitatively similar to those observed within $-3/2 < D/|J| < 0$ and, therefore, in the following we will limit our considerations to the values $-1 \leq D/|J| \leq 1$.

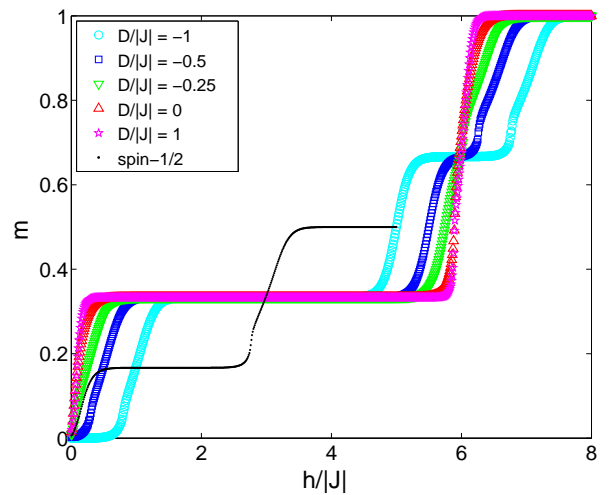

**Fig. 2.** Magnetization as a function of the reduced external field for selected values of the reduced single-ion anisotropy and reduced temperature $k_B T/|J| = 0.1$. The dotted curve corresponds to the spin-1/2 case (equivalent to the spin-1 case with $D/|J| \to \infty$).

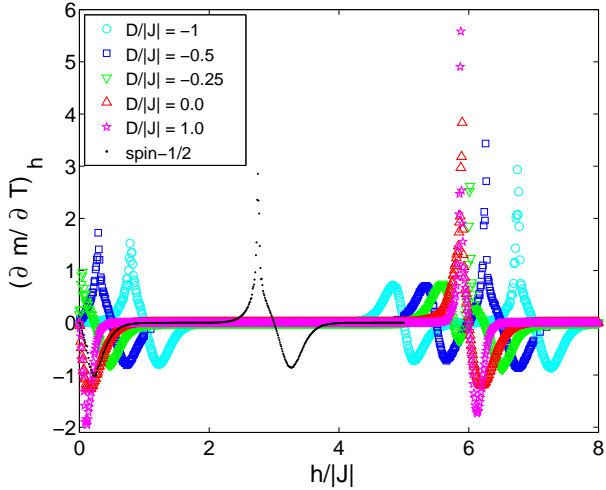

**Fig. 3.** Field-dependence of the derivative of magnetization with respect to temperature at constant field for $k_B T/|J| = 0.1$.

In figure 3 we present responses of the magnetizations to the change of temperature at a constant field. As expected, the most prominent changes are observed close to the field-induced phase transitions, while almost no changes can be seen within the respective plateaus. However, the responses are again different for $D/|J| \geq 0$ from those at $D/|J| < 0$. Typically, the onset of a plateau in the field-increasing process is accompanied with initial period of a positive magnetization change followed by an interval of a negative change. However, this is not the case for $D/|J| \geq 0$ at the onset of the first (ferrimagnetic (↑↓↑)) plateau, where the magnetization change is only negative.

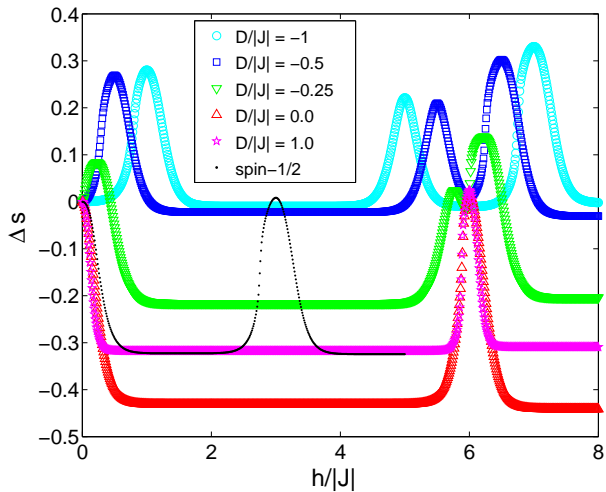

**Fig. 4.** Field-dependence of the isothermal entropy change for $k_B T/|J| = 0.1$.

This behavior of the magnetization responses to the change of temperature translates into the magnetocaloric properties, as evidenced in figure 4. It is clear that for $D/|J| \geq 0$ the initial period of the negative magnetization change must result in a negative entropy change. The latter remains negative until it is compensated by a period of a positive magnetization change at the beginning of the onset of the saturated (↑↑↑) phase, followed by another plunge into negative values. We note that this scenario applies also to the spin-1/2 TLIA model, in spite of the claim in Ref. [9] that the entropy changes sign as the field is increased. Comparison of the results in Ref. [9] for $\Delta s$ with our results for $(\partial m/\partial T)_h$ makes us believe that the authors mistook the two quantities and thus their claim refers to the behavior of $(\partial m/\partial T)_h$ instead of $\Delta s$. On the other hand, the spin-1/2 entropy change curve, shown in figure 4, indicates that $\Delta s$ may indeed shortly switch to small positive values at the onset to the saturation magnetization plateau. Such a scenario cannot be ruled out but we think that it is more likely just a result of a numerical error, in particular the one associated with problematic numerical quadrature of the spike-like function $(\partial m/\partial T)_h$. Nevertheless, changing between positive and negative values of $\Delta s$ is evident for small $D/|J| < 0$, such as $D/|J| = -0.25$. Finally, for larger negative $D/|J|$ the entropy changes appear to remain non-negative for any $h/|J|$. Again, it is not clear whether the small negative values obtained within the plateaus are real or just artifacts of numerical quadrature errors.

## 4 Conclusions

We studied low-temperature magnetization processes and magnetocaloric properties of a geometrically frustrated spin-1 Blume-Capel model on a triangular lattice. We found that the present model may display qualitatively different behavior, depending on the sign of the single-ion anisotropy. For positive values the behavior is similar to what we can observe in the spin-1/2 Ising antiferromagnet on a triangular lattice. Namely, in the field-increasing processes before the saturation state there is a magnetization plateau with the height of 1/3 of the saturation value and the isothermal entropy change appears to remain negative at any increase of the field, thus showing the direct magnetocaloric effect (MCE). It is worthwhile noticing that such a behavior is untypical for antiferromagnets and can be ascribed to the presence of geometrical frustration.. More specifically, in the ordered phase at low temperatures in a field regular nonfrustrated antiferromagnets are always expected to show positive magnetization change with increasing temperature and thus IMCE [3]. On the other hand, the magnetization of the spin-1/2 TLIA model with increasing temperature can either decrease or increase, depending on the field value (see e.g. [10]). For negative values of the single-ion anisotropy, as the field increases before the saturation value is reached the system consecutively passes through three phases associated with three magnetization plateaus with the heights of 0, 1/3 and 2/3 of the saturation value. For small negative values of the single-ion anisotropy the entropy change can be either negative (the direct MCE) or positive (the inverse MCE), depending on the field intensity, and for larger negative values the entropy change is positive for any field value.

The present investigations were performed at a fixed low temperature. However, for the compound to be used as practical magnetic refrigerant it is important that the magnetic entropy change persists over certain temperature range. Therefore, our further effort will focus on the study of temperature dependences of the observed magnetocaloric effects.

## Acknowledgements

This work was supported by the Scientific Grant Agency of Ministry of Education of Slovak Republic (Grant No. 1/0234/12). The authors acknowledge the financial support by the ERDF EU (European Union European Regional Development Fund) grant provided under the contract No. ITMS26220120047 (activity 3.2.).